# Effective Lifetime of Non-Equilibrium Carriers in Semiconductors from Non-Adiabatic Molecular Dynamics Simulations

Shanshan Wang[1,2], Menglin Huang[1], Yu-Ning Wu[2], Weibin Chu[1], Jin Zhao[3], Aron Walsh[4], Xin-Gao Gong[1,5], Su-Huai Wei[6], and Shiyou Chen[1,2,5,*]

[1]Key Laboratory of Computational Physical Sciences (MOE), and State Key Laboratory of ASIC and System, School of Microelectronics, Fudan University, Shanghai 200433, China, *chensy@fudan.edu.cn
[2]Key Laboratory of Polar Materials and Devices (MOE) and Department of Electronics, East China Normal University, Shanghai 200241, China
[3]ICQD/Hefei National Laboratory for Physical Sciences at the Microscale, CAS Key Laboratory of Strongly-Coupled Quantum Matter Physics, and Department of Physics, University of Science and Technology of China, Hefei, Anhui 230026, China
[4]Department of Materials, Imperial College London, Exhibition Road, London SW7 2AZ, UK
[5]Shanghai Qi Zhi Institute, Shanghai 200030, China
[6]Beijing Computational Science Research Center, Beijing 100193, China

The lifetime of non-equilibrium electrons and holes in semiconductors is crucial for solar cell and optoelectronic applications. Non-adiabatic molecular dynamics (NAMD) simulations based on time-dependent density functional theory (TDDFT) are widely used to study excited-state carrier dynamics. However, the calculated carrier lifetimes are often different from experimental results by orders of magnitude. In this work, by revisiting the definition of carrier lifetime and considering different recombination mechanisms, we report a systematic procedure for calculating the effective carrier lifetime in realistic semiconductor crystals that can be compared directly to experimental measurements. The procedure shows that considering all recombination mechanisms and using reasonable densities of carriers and defects are crucial in calculating the effective lifetime. When NAMD simulations consider only Shockey-Read-Hall (SRH) defect-assisted and band-to-band non-radiative recombination while neglect band-to-band radiative recombination, and the densities of non-equilibrium carriers and defects in supercell simulations are much higher than those in realistic semiconductors under solar illumination, the calculated lifetimes are ineffective and thus differ from experiments. Using our procedure, the calculated effective lifetime of the halide perovskite $CH_3NH_3PbI_3$ agrees with experiments. It is mainly determined by band-to-band radiative and defect-assisted non-radiative recombination, while band-to-band non-radiative recombination is negligible. These results indicate that it is possible to calculate carrier lifetimes accurately based on NAMD simulations, but the directly calculated values should be converted to effective lifetimes for comparison to experiments. The revised procedure can be widely applied in future carrier lifetime simulations.

Excited-state carrier dynamics in semiconductors is fundamental to the development of optoelectronic[1], photovoltaic[2], photocatalytic[3] and other functional devices working under light illumination. The lifetime of the photo-excited non-equilibrium carriers is an important quantity determining the performance of these devices, *e.g.*, it determines the diffusion length of the minority carriers in photovoltaic semiconductors and is thus critical to the efficiency of solar cells. For decades, this important quantity is obtained mainly through the ultrafast time-resolved photoluminescence spectroscopy[4] and transient absorption spectroscopy[5,6].

Theoretically, conventional first-principles molecular dynamics based on adiabatic approximation does not describe the excited-state carrier dynamics. However, rapid developments in nonadiabatic molecular dynamics (NAMD)[7,8] based on time-dependent DFT (TDDFT)[9] methods in the past decade has made the simulation of excited-state carrier dynamics in semiconductors possible. A series of breakthroughs have been reported recently[10-20]. Therefore, the first-principles prediction of carrier lifetime is attracting wide attention. There are many groups worldwide who have adopted the NAMD simulations based on Erhenfest or surface hopping schemes to predict the lifetime of non-equilibrium carriers[21-42]. In these studies, non-equilibrium carrier populations were produced by exciting an electron from the occupied level to a higher-energy unoccupied level in the supercell calculations, *e.g.*, from the valence band maximum (VBM) level to the conduction band minimum (CBM) level. Then NAMD simulations were performed and the lifetime $\tau$ was calculated through fitting the decay of the electron population on the excited-state level to the function $\Delta n(t) \propto \exp(-t/\tau)$ [21-42]. For example, the NAMD simulation of Qiao *et al.* predicted that the lifetime of photo-excited carriers is 1.5 ns in $CH_3NH_3PbI_3$[43], and similar values were reported in other studies[44-47]. However, plenty of experiments have reported that the actual lifetime of photo-excited carriers can be as long as several microseconds in $CH_3NH_3PbI_3$[48-52]. There is a discrepancy between the calculated and experimental values.

To reveal the origin of the discrepancy, we repeated the NAMD simulation using the procedures as in Ref. [43] and found that our result agrees with theirs, so technical errors in the simulations can be excluded. Kim and Walsh pointed out that non-adiabatic coupling matrix elements between the valence and conduction bands may be significantly overestimated in the NAMD simulations, which results in exaggerated non-radiative recombination in pristine CH$_3$NH$_3$PbI$_3$ and thus causes the short lifetime[53]. That means the accuracy issues of the current NAMD methods may be one possible origin of the discrepancy. On the other hand, as noted by Qiao *et al.*[43] and Chu *et al.*[54], most of the present NAMD simulations used supercells with a limited size, so the simulations assumed a certain carrier density, defect density, and light intensity, which might be very different from the conditions in real samples and can thus also be the origin of the discrepancy. If this is the origin, then two open questions appear: (i) what is the meaning of the directly calculated carrier lifetime from the NAMD simulations; (ii) is it possible to use the NAMD simulations to calculate the carrier lifetime that can be compared directly to the experimentally measured lifetime of the real semiconductor samples with the actual carrier density and defect density, and working under solar illumination?

In this work, we started from the fundamental definition of the lifetime of non-equilibrium carriers and developed a systematic procedure for calculating the effective carrier lifetime in real samples based on the results from the NAMD simulations and other calculations. The calculated effective carrier lifetime can be compared directly to experimental values. Using this procedure, we find that the reported lifetime (1.5 ns) of CH$_3$NH$_3$PbI$_3$ in previous NAMD studies is the seriously underestimated lifetime $\tau_{band-band}^{non-rad}$ of band-to-band non-radiative recombination, and the effective $\tau_{band-band}^{non-rad}$ is actually very long (150 μs). The effective carrier lifetime in CH$_3$NH$_3$PbI$_3$ is mainly determined by the band-to-band radiative and defect-assisted SRH recombination, while the influence of band-to-band non-radiative recombination is negligible. When all recombination mechanisms are considered and the effects are summed using the systematic procedure, the effective lifetime is consistent with the experimental lifetime and can explain the observed efficient photoluminescence of CH$_3$NH$_3$PbI$_3$. Besides that, we also revisited the carrier lifetime of three other optoelectronic semiconductors reported in recent NAMD studies and found that the conclusions can be changed if all the recombination mechanisms are considered and reasonable carrier and defect densities are used in the simulation. We propose that it is necessary to calculate the effective lifetime using the systematic procedure outlined here in future carrier dynamics studies to provide the correct understanding of the mechanisms that determine the experimentally measured carrier lifetime.

**Definition of Carrier Lifetime and Systematic Calculation Procedure.**

As discussed in many textbooks of semiconductor physics[55], the time derivative of the non-equilibrium carrier density $\Delta n(t)$ is equal to the difference between the generation rate $G$ and the recombination rate $U$ of non-equilibrium carriers,

$$\frac{d\Delta n(t)}{dt} = G(t) - U(t) \quad (1).$$

When the generation stops, $G(t)=0$, $\Delta n(t)$ will decay with the rate $-U(t)$. If the recombination rate $U(t)$ depends linearly on $\Delta n(t)$, i.e.,

$$U(t) = \frac{\Delta n(t)}{\tau} \quad (2)$$

where $\tau$ is a constant,

$$\tau = \frac{\Delta n(t)}{U(t)} \quad (3),$$

then the decay of $\Delta n(t)$ will follow,

$$\Delta n(t) = \Delta n(0)\exp\left(-\frac{U(t)}{\Delta n(t)}t\right) = \Delta n(0)\exp\left(-\frac{t}{\tau}\right) \quad (4)$$

where $\Delta n(0)$ is the density of non-equilibrium carriers at the $t = 0$ moment when the generation stops. As we can see, in this case, the constant $\tau$ means the time when the density decays to *1/e* of the original value $\Delta n(0)$ after the generation stops, so it is defined as the lifetime of non-equilibrium carriers. Then, Eq. (1) becomes,

$$\frac{d\Delta n(t)}{dt} = -U(t) = -\frac{\Delta n(t)}{\tau} \quad (5).$$

In standard NAMD studies, the lifetime $\tau$ is calculated through fitting the decay of the electron population on the excited-state level to the function $P(t) \propto \exp(-t/\tau)$, which originates from the decay function Eq. (4).

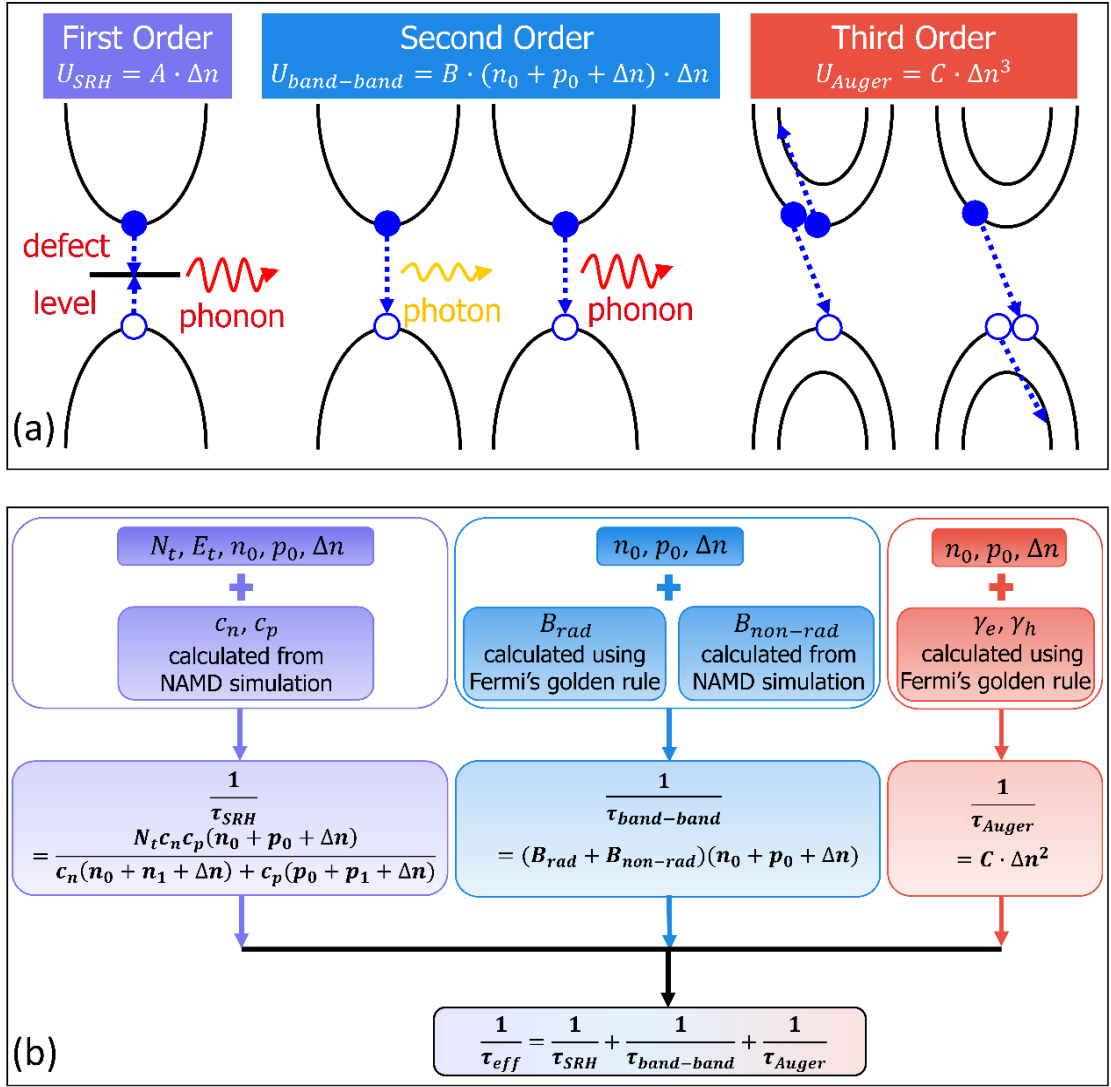

Figure 1. The schematic plot of the first-order defect-assisted SRH, the second-order band-to-band radiative or non-radiative, and the third-order Auger recombination (a), and the procedure for calculating the effective lifetime with the three major recombination mechanisms considered (b).

According to Eq. (3), the lifetime $\tau$ is determined by the ratio between the density of non-equilibrium carriers $\Delta n(t)$ and the recombination rate $U(t)$. In semiconductors, as shown in Fig. 1(a), there are many possible recombination mechanisms, *e.g.*, the defect-assisted non-radiative SRH recombination, the band-to-band recombination, and the Auger recombination. Therefore, the recombination rate $U(t)$ is the sum of the rates of all these mechanisms,

$$U(t) = U_{SRH}(t) + U_{band-band}(t) + U_{Auger}(t) \quad (6).$$

Correspondingly, the lifetime that is effective in real semiconductor samples should also be contributed by all these mechanisms,

$$\frac{1}{\tau} = \frac{1}{\tau_{SRH}} + \frac{1}{\tau_{band-band}} + \frac{1}{\tau_{Auger}} \quad (7)$$

where $\tau_{SRH}$, $\tau_{band-band}$ and $\tau_{Auger}$ are the lifetime when only one of the three mechanisms are considered. With this definition, we can also write,

$$\tau_{SRH} = \frac{\Delta n(t)}{U_{SRH}} \quad (8),$$

$$\tau_{band-band} = \frac{\Delta n(t)}{U_{band-band}} \quad (9),$$

$$\tau_{Auger} = \frac{\Delta n(t)}{U_{Auger}} \quad (10).$$

As shown in Refs. [55] for photo-excited non-equilibrium carriers (the densities of non-equilibrium electrons and holes are equal, $\Delta n(t) = \Delta p(t)$ ), the three recombination rates depend not only on $\Delta n(t)$, but also on the electron carrier density $n_0$ and hole carrier density $p_0$ under the equilibrium state, as described by,

$$U_{SRH}(t) = A \cdot \Delta n(t) \quad (11),$$

$$U_{band-band}(t) = \Delta n(t) \cdot B \cdot [n_0 + p_0 + \Delta n(t)] \quad (12),$$

$$U_{Auger}(t) = C \cdot \Delta n(t)^3 \quad (13)$$

in which, *A*, *B* and *C* are the SRH defect-assisted non-radiative, band-to-band and Auger recombination coefficients, respectively; $n_0$ and $p_0$ can be calculated as $n_0 = N_c exp(\frac{E_F - E_{CBM}}{k_0 T})$ and $p_0 = N_v \exp(\frac{E_{VBM} - E_F}{k_0 T})$, where T is the temperature, $k_0$ is the Boltzmann constant, $E_F$ is the Fermi level, $E_{VBM}$ is the valence band maximum level, $E_{CBM}$ is the conduction band minimum level, and $N_v$ and $N_c$ are the effective density of states for valence band and conduction band edges, respectively.

The SRH recombination coefficient *A* can be calculated following[56-58],

$$A = \frac{N_t c_n c_p (n_0 + p_0 + \Delta n(t))}{c_n(n_0 + n_1 + \Delta n(t)) + c_p(p_0 + p_1 + \Delta n(t))} \quad (14)$$

where $N_t$ is the density of the recombination-center defects; $c_n$ and $c_p$ are the electron capture coefficient and hole capture coefficient of the defect level, respectively; $n_1 =$

$N_c exp(\frac{E_t-E_{CBM}}{k_0 T})$, $p_1 = N_v exp(\frac{E_{VBM}-E_t}{k_0 T})$ in which $E_t$ is the energy of defect level.

The Auger recombination coefficient $C$ can be calculated following[55],

$$C = \frac{(\gamma_e n_0 + \gamma_h p_0)[n_0 + p_0 + \Delta n(t)] + (\gamma_e + \gamma_h)[n_0 + p_0 + \Delta n(t)]\Delta n(t)}{\Delta n(t)^2} \quad (15)$$

in which $\gamma_e$ and $\gamma_h$ represent the *e-e-h* Auger recombination coefficient (one electron at the conduction band is excited to higher-level state when one electron-hole pair recombines) and *h-h-e* Auger recombination coefficient (one hole at the valence band is promoted to lower-level state when one electron-hole pair recombines).

According to Eqs. (11-15), we can notice that the recombination rate *U(t)* (including $U_{SRH}(t)$, $U_{band-band}(t)$ and $U_{Auger}(t)$) depends non-linearly on $\Delta n(t)$, so $\tau$ (the carrier lifetime) is not a constant with respect to $\Delta n(t)$. Therefore, in principle, the decay of $\Delta n(t)$ does not follow the simple function in Eq. (4). When $\tau$ was defined as the carrier lifetime (the time when the density decays to *1/e* of the original value $\Delta n(0)$ after the generation stops) and was calculated through fitting the decay of $\Delta n(t)$ to the function $\exp(-t/\tau)$, we assumed implicitly that $\tau$ is a constant and *U(t)* depends linearly on $\Delta n(t)$ in the following recombination processes after the generation stops. Of course, the fitting cannot be good if $\tau$ has a strong dependence on $\Delta n(t)$. Therefore, when we discuss the carrier lifetime $\tau$, we must pay attention to this dependence and note that $\tau$ can be approximated as a constant only for a small range of *t* in which $\Delta n(t)$ does not change significantly. In such a small range of *t*, $\Delta n(t)$ can be approximated as $\Delta n(0)$ which is the value at the moment when the generation stops. With $\Delta n(0)$, the recombination coefficients *A* and *B* can thus be determined, then the corresponding lifetime can be calculated as,

$$\tau_{SRH} = \frac{1}{A} \quad (16)$$

$$\tau_{band-band} = \frac{1}{B \cdot [n_0 + p_0 + \Delta n(0)]} \quad (17)$$

$$\tau_{Auger} = \frac{1}{(\gamma_e n_0 + \gamma_h p_0)[n_0 + p_0 + \Delta n(0)] + (\gamma_e + \gamma_h)[n_0 + p_0 + \Delta n(0)]\Delta n(0)} \quad (18).$$

In operating devices under continuous illumination, the density of non-equilibrium carriers should reach a steady state after the generation and recombination are counterbalanced, then the steady-state density should be considered as $\Delta n(0)$ when calculating the carrier lifetime according to Eqs. (16), (17) and (18).

As shown in Fig. 1a, the defect-assisted SRH non-radiative recombination occurs between one carrier (electron or hole) and one defect, so the dependence of $U_{SRH}$ on $\Delta n$ (abbreviation of $\Delta n(0)$) is first order in Eq. (11); the band-to-band recombination occurs between two carriers (one electron and one hole), so the dependence of $U_{band-band}$ on $\Delta n$ is second order in Eq. (12); the Auger recombination occurs between three carriers (two electrons and one hole, or one electron and two holes), so the dependence of $U_{Auger}$ on $\Delta n$ is third order in Eq. (13). Therefore, when the density $\Delta n$ of photo-excited non-equilibrium carriers is small, $U_{SRH}$ and $U_{band-band}$ are much larger than $U_{Auger}$, thus $\tau_{SRH}$ and $\tau_{band-band}$ determine $\tau$; when $\Delta n$ is large, $U_{SRH}$ and $U_{band-band}$ are much smaller than $U_{Auger}$, thus $\tau_{Auger}$ determines $\tau$[59].

In order to calculate the total effective lifetime $\tau$ of non-equilibrium carriers in real semiconductor samples with a certain defect density $N_t$, equilibrium carrier density $n_0$ ($p_0$), and non-equilibrium carrier density $\Delta n(0)$, $\tau_{SRH}$, $\tau_{band-band}$ and $\tau_{Auger}$ should all be calculated and combined. A procedure is plotted in Fig. 1b to show the flow of the calculations, including:

(1) For a given type of defect with the energy level $E_t$, calculate its electron and hole capture coefficients $c_n$, $c_p$. They can be calculated from the NAMD simulation, as discussed in the following section "Dependence of $\tau_{SRH}$ on Density of Recombination-Center Defects", or using the non-radiative multi-phonon theory[60,61]. With $E_t$, $c_n$ and $c_p$, $\tau_{SRH}$ can be calculated for given $N_t$, $n_0$ ($p_0$) and $\Delta n(0)$.

(2) Calculate the band-to-band recombination coefficient $B$. The detailed methods will be discussed in the following sections "Non-Radiative Carrier Recombination in NAMD Simulations" and "Dependence of $\tau_{band-band}$ on Density of Non-Equilibrium Carriers". With $B$, $\tau_{band-band}$ can be calculated for given $n_0$ ($p_0$) and $\Delta n(0)$.

(3) Calculate the Auger recombination coefficients $\gamma_e$ and $\gamma_h$. The methods have been discussed in Ref. [62]. With $\gamma_e$ and $\gamma_h$, $\tau_{Auger}$ can be calculated for given $n_0$ ($p_0$) and $\Delta n(0)$.

After the calculation of $\tau_{SRH}$, $\tau_{band-band}$ and $\tau_{Auger}$, the effective lifetime $\tau$ can be derived according to Eq. (7) and the value can be compared directly to the experimentally measured lifetime in real samples with the given $N_t$, $n_0$ ($p_0$) and $\Delta n(0)$.

**Non-Radiative Recombination in NAMD Simulations.**

With the formulae derived for calculating the effective carrier lifetime, now we can analyze why the directly calculated lifetime from the recent NAMD simulations differ from experimental values for CH$_3$NH$_3$PbI$_3$.

These simulations adopted supercells with several hundreds of atoms, and considered the cases for both the defect-free supercells and those with defects. For the defect-free supercell, there are no electronic states in the band gap and the recombination occurs between the electron on CBM and the hole on VBM, so the band-to-band recombination is simulated and the derived lifetime is $\tau_{band-band}$. For the supercell with a defect, if a defect level is produced in the band gap, the defect-assisted SRH non-radiative recombination is simulated and the derived lifetime is $\tau_{SRH}$. For both cases, the simulated lifetime ($\tau_{band-band}$ or $\tau_{SRH}$) is only a part of the effective lifetime $\tau$ as described by Eq. (7). Incomplete consideration of the recombination mechanisms may cause a large difference between the simulated and experimental results. More importantly, we should pay attention to the dependence of the results on $\Delta n(0)$ and $N_t$ assumed in the simulations. In most simulations, the non-equilibrium electron and hole carriers are generated through exciting an electron from the VBM to CBM in the supercell. If the supercell has several hundred atoms, the density $\Delta n(0)$ of the non-equilibrium carriers is as high as $10^{20}$ cm$^{-3}$ (for CH$_3$NH$_3$PbI$_3$ supercell with 192 atoms, the density is $2.6 \times 10^{20}$ cm$^{-3}$). However, under the normal sunlight illumination, the excited non-equilibrium electron carriers have a much lower density, only around $10^{13}$-$10^{15}$ cm$^{-3}$ (as calculated below). Obviously, the assumed $\Delta n(0)$ (similarly for the defect density $N_t$) in the NAMD supercell simulations is much higher than that of the density in real devices working under sunlight illumination, as compared in Fig. 2. Eqs. (16) and (17) show that $\tau_{band-band}$ and $\tau_{SRH}$ depend on $\Delta n(0)$ and $N_t$, so the simulated lifetime actually is the result at the very high $\Delta n(0)$ and $N_t$, rather than the lifetime in real samples with lower $\Delta n(0)$ and $N_t$, which is another origin of the large difference.

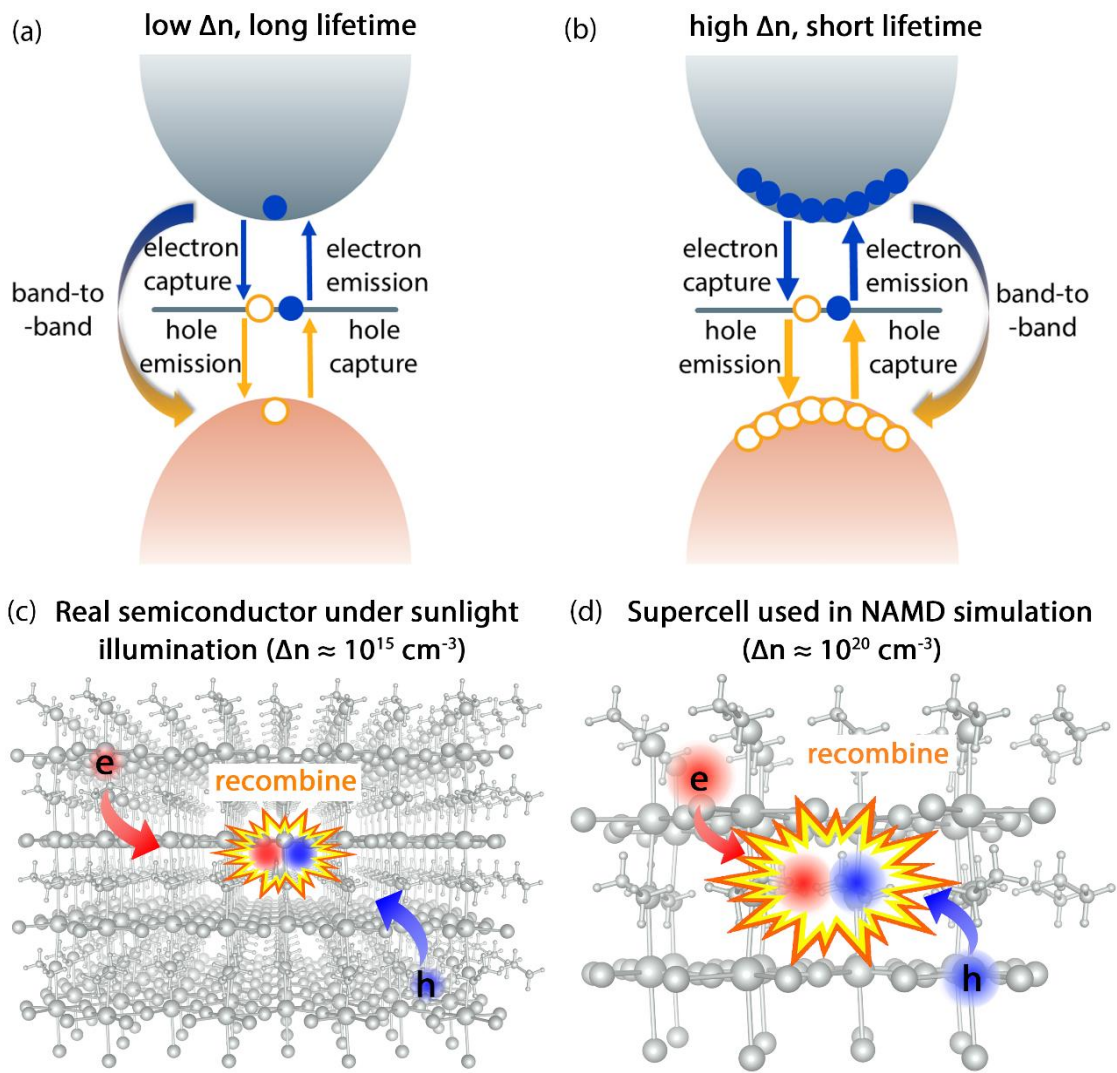

Figure 2. Schematic plot of band-to-band and defect-assisted (SRH) recombination processes in semiconductors with low and high densities of non-equilibrium carriers: (a) and (b) in the reciprocal band-structure space with electron carriers on the conduction band and hole carriers on the valence band; (c) and (d) in the real semiconductor lattice space. The normal density of non-equilibrium carriers is usually around $10^{15}$ cm$^{-3}$ under sunlight illumination, as shown in (a) and (c), while the density of the non-equilibrium carriers produced in the several-hundred-atom supercell for NAMD simulations can exceed $10^{20}$ cm$^{-3}$, as shown in (b) and (d), which causes fast electron-hole recombination.

Attention should also be paid to the meaning of the simulated $\tau_{band-band}$. As shown in Fig. 1a, there are two types of band-to-band recombination: radiative and non-radiative. For band-to-band radiative recombination, the energy of non-equilibrium carriers is converted into the energy of an emitted photon. For band-to-band non-radiative recombination, the energy of non-equilibrium carriers is converted into the energy of the vibration energy of ions through electron-phonon interactions. The recombination rate and lifetime are contributed by both types, as described by,

$$U_{band-band}(t) = U_{band-band}^{rad}(t) + U_{band-band}^{non-rad}(t)$$

$$= \Delta n(t) \cdot (B_{rad} + B_{non-rad}) \cdot [n_0 + p_0 + \Delta n(t)] \quad (19)$$

$$\frac{1}{\tau_{band-band}} = \frac{1}{\tau_{band-band}^{rad}} + \frac{1}{\tau_{band-band}^{non-rad}} \quad (20).$$

Correspondingly, the recombination coefficient $B$ is also contributed by two parts, $B_{rad}$ and $B_{non-rad}$.

In the standard textbook description, band-to-band recombination is considered as a radiative process,[63] while the non-radiative rate is low under normal operation conditions. The corresponding recombination rate $U_{band-band}^{rad}$ and coefficient $B_{rad}$ can be calculated using the Fermi's golden rule and the transition dipole moment (momentum matrix elements)[63-66].

However, in recent NAMD simulations[21-25,31-35,37-40] an excitation is modelled through a change in occupation numbers of valence and conduction bands of the supercell. Recombination towards the ground state is then simulated by NAMD and the band-to-band recombination lifetime $\tau$ is calculated by fitting the decay of the electron population using $P(t) = \exp(-t/\tau)$. These simulations are in the microcanonical (NVE) ensemble, so the electronic energy decrease is converted into vibrational (kinetic) energy through the electron-phonon coupling, which obeys the energy conservation rule. Emission of light is not considered and the associated lifetime is in fact $\tau_{band-band}^{non-rad}$. This was also pointed out by Kim and Walsh[53].

The contribution $\tau_{band-band}^{non-rad}$ is just part of the $\tau_{band-band}$, whereas $\tau_{band-band}$ is just part of the true effective $\tau$, as shown in Eq. (7). Therefore, the meaning of $\tau_{band-band}^{non-rad}$ calculated in the NAMD studies is different from the effective lifetime that can be measured experimentally. Only when the lifetimes for other mechanisms, such as band-to-band radiative $\tau_{band-band}^{rad}$ and defect-assisted non-radiative $\tau_{SRH}$, are longer, can the calculated $\tau_{band-band}^{non-rad}$ be important. Unfortunately, the importance of the band-to-band non-radiative recombination may be overestimated due to the high carrier density in small supercell models, as discussed next.

**Dependence of $\tau_{band-band}$ on Density of Non-Equilibrium Carriers.**

Eq. (17) shows that $\tau_{band-band}$ (similarly for $\tau_{band-band}^{non-rad}$) depends not only on the density of non-equilibrium carriers $\Delta n(0)$, but also on the density of equilibrium carriers, $n_0$ and $p_0$. If the recombination coefficient $B_{non-rad}$ is known, the dependence of $\tau_{band-band}^{non-rad}$ on $\Delta n(0)$, $n_0$ and $p_0$ can be directly calculated. An example is given in Fig. 3a and Fig. 3b, where we assume $B_{non-rad}=2.6 \times 10^{-12}$ cm³/s.

Fig. 3a shows the decrease of $\tau_{band-band}^{non-rad}$ as $\Delta n(0)$ increases from $10^{10}$ cm⁻³ to $10^{20}$ cm⁻³ for the given $n_0$ and $p_0$. Obviously, $\tau_{band-band}^{non-rad}$ can be as long as hundreds of microseconds if $\Delta n(0)$ is on the order of $10^{15}$ cm⁻³, while the value decreases quickly to several nanoseconds if $\Delta n(0)$ increases to $10^{20}$ cm⁻³. The large difference can be seen more directly in Fig. 3c and 3d, which show the simulated decay of the non-equilibrium carrier density $\Delta n(t)$ for different initial density $\Delta n(0)$ according to $\frac{d\Delta n(t)}{dt} = -U = -U_{band-band}^{non-rad}$. To demonstrate the influence of the initial density $\Delta n(0)$, $\Delta n(0)$ is set to $2.6 \times 10^{20}$ cm⁻³ in Fig. 3c and $2.6 \times 10^{15}$ cm⁻³ in Fig. 3d, which has a difference as large as 5 orders of magnitude. As we can see, the decay in Fig. 3c is very fast and it takes only 1-2 ns for $\Delta n(t)$ decaying to 1/e of $\Delta n(0)$, consistent with the short $\tau_{band-band}^{non-rad}=1.5$ ns at $\Delta n(0)=2.6 \times 10^{20}$ cm⁻³ in Fig. 3a (for $n_0=p_0=10^6$ cm⁻³). In contrast, the decay in Fig. 3d is slower and it takes almost 0.1 ms for $\Delta n(t)$ decaying to 1/e of $\Delta n(0)$, consistent with the long $\tau_{band-band}^{non-rad}=0.15$ ms according to Eq. (17). The shorter $\tau_{band-band}^{non-rad}$ for higher $\Delta n(0)$ is easy to understand. As shown schematically in Fig. 2c, when $\Delta n(0)$ is high, there are more non-equilibrium electron and hole carriers in the structure, so they have more chances to interact and recombine. In contrast, when $\Delta n(0)$ is low in Fig. 2d, there are fewer chances for them to interact and the recombination rate is reduced.

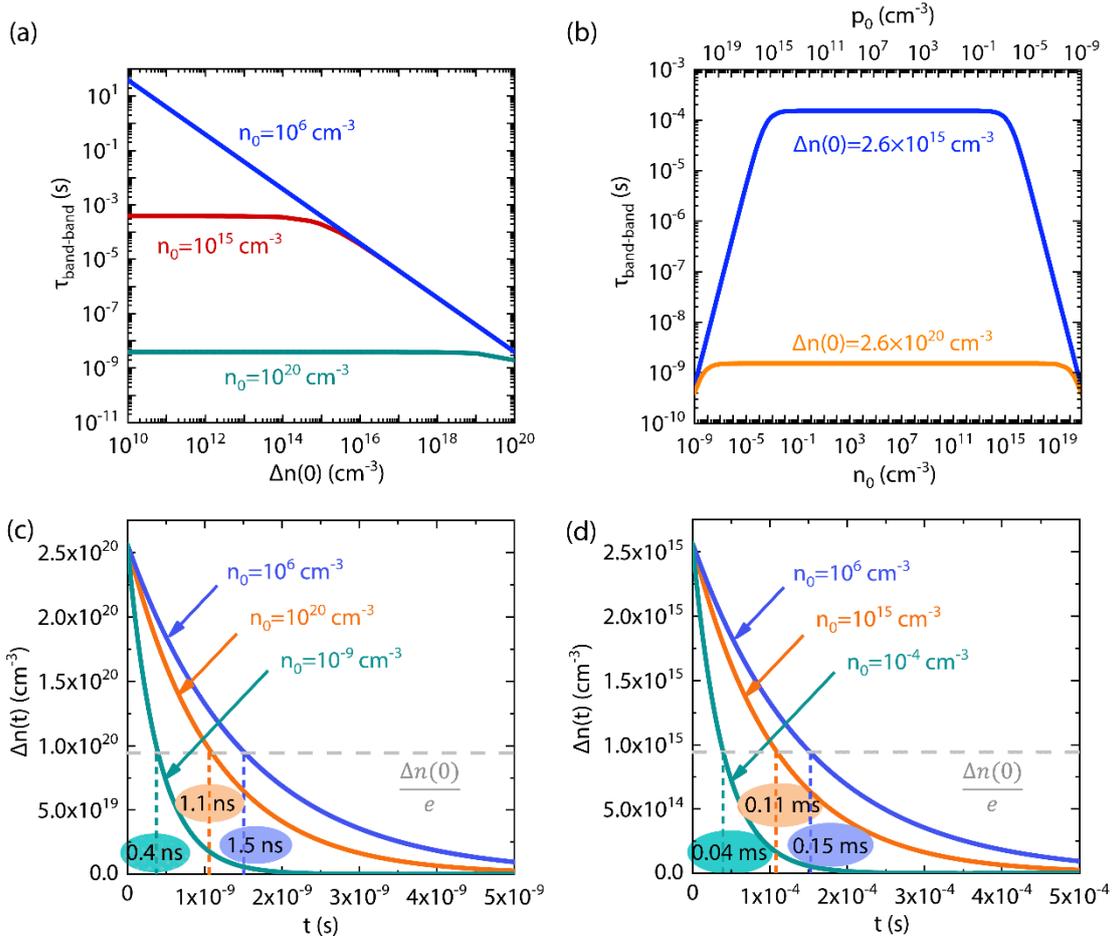

Figure 3. Calculated dependence of $\tau_{band-band}^{non-rad}$ on (a) the density of non-equilibrium carriers $\Delta n(0)$ for the given density of equilibrium carriers $n_0$ ($p_0$), and (b) $n_0$ ($p_0$) for the given $\Delta n(0)$; and the time evolution of $\Delta n(t)$ for (c) high initial density $\Delta n(0)=2.6 \times 10^{20}$ cm$^{-3}$ and (d) low $\Delta n(0)=2.6 \times 10^{15}$ cm$^{-3}$, when only band-to-band nonradiative recombination is considered. In (c) and (d), three different sets of $n_0$ ($p_0$) are considered.

The short $\tau_{band-band}^{non-rad}$ at high $\Delta n(0)$ can be used to explain why the NAMD simulated lifetime is only 1.5 ns in CH$_3$NH$_3$PbI$_3$[43], much shorter than the experimental lifetime of several microseconds. As compared in Fig. 2, $\Delta n(0)$ assumed in the NAMD simulations is high (around $10^{20}$ cm$^{-3}$), much higher than typical values (around $10^{15}$ cm$^{-3}$) in semiconductor samples under one-sun illumination. If instead, the simulation uses a very large supercell with $10^7$ atoms, $\Delta n(0)$ decreases to around $10^{15}$ cm$^{-3}$, then $\tau_{band-band}^{non-rad}$ can be as long as $10^5$ ns, as shown in Fig. 3d. Based on the analysis, the large difference between the recent simulations and experiments for CH$_3$NH$_3$PbI$_3$ can now be understood, *i.e.*, the short lifetime of 1.5 ns from NAMD simulations is not the effective $\tau_{band-band}^{non-rad}$ or the total effective $\tau$ in real samples, so it is meaningless to compare the value to the experimental lifetime directly. Only after all the contributions of the effective $\tau_{band-band}^{non-rad}$, $\tau_{band-band}^{rad}$, $\tau_{SRH}$ and $\tau_{Auger}$ are considered at the actual $\Delta n(0)$ and the total effective $\tau$ is calculated, can comparison to measurements be made.

A suitable way to predict the effective $\tau_{band-band}^{non-rad}$ is by adopting Eq. (17) to derive the supercell-size-independent recombination coefficient $B_{non-rad}$ from the raw $\tau_{band-band}^{non-rad}$ derived from the NAMD supercell simulations. Eq. (17) is changed into

$$B_{non-rad} = \frac{1}{\tau_{band-band}^{non-rad}[n_0+p_0+\Delta n(0)]} \quad (21).$$

The densities of equilibrium carriers, $n_0$ and $p_0$, should be small in the defect-free (dopant-free) supercell and can be neglected compared to $\Delta n(0)$, because the band gap is above 1.5 eV and the thermal excitation of carriers can be neglected. As a result, $B_{non-rad} = \frac{1}{\tau_{band-band}^{non-rad}*\Delta n(0)}$.

To take one example, Qiao *et al.* derived a lifetime $\tau_{band-band}^{non-rad}$ of 1.5 ns from the NAMD simulation using a 192-atom CH$_3$NH$_3$PbI$_3$ supercell[43], in which $\Delta n(0)$=2.6 × 10$^{20}$ cm$^{-3}$. Then the derived $B_{non-rad} \approx 2.6 \times 10^{-12}$ cm$^3$/s. With the derived $B_{non-rad}$, we can calculate the effective $\tau_{band-band}^{non-rad}$ for low $\Delta n(0)$ according to Eq. (17). For standard $\Delta n(0)$ around 10$^{15}$ cm$^{-3}$ under sunlight illumination, the effective $\tau_{band-band}^{non-rad}$ should be around 150 μs, which is higher than the directly derived value of 1.5 ns by 5 orders of magnitude. As a result, the recombination coefficient $B_{non-rad}$ derived from the NAMD simulations[43] gives a long $\tau_{band-band}^{non-rad}$ when $\Delta n(0)$ is set to a reasonable value. Such a long $\tau_{band-band}^{non-rad}$ is in accordance with longstanding non-radiative carrier capture theories (which stem from Landau-Zener theory that describes the probability of transition following a non-adiabatic level crossing, however, the band gap of CH$_3$NH$_3$PbI$_3$ acts as a substantial barrier that prohibits band-to-band non-radiative recombination)[53]. If the effective $\tau_{SRH}$ or $\tau_{band-band}^{rad}$ is shorter than 10 μs, the contribution of the long effective $\tau_{band-band}^{non-rad}$ (150 μs) to the total effective $\tau$ can be negligible. The comparison of $\tau_{SRH}$, $\tau_{band-band}^{rad}$ and $\tau_{band-band}^{non-rad}$ will be discussed later.

With the approach to determine the effective $\tau_{band-band}^{non-rad}$, we can revisit other NAMD studies which reported carrier lifetimes. Ghosh *et al.* investigated the excited-state carrier dynamics near the band edges of (BA)$_2$PbBr$_4$ and (PEA)$_2$PbBr$_4$ perovskites using the NAMD simulation and the calculated lifetime is 668 ps and 2158 ps, respectively, for $\Delta n(0)=10^{20}$ cm$^{-3}$ in the supercell[31]. Therefore, the corresponding coefficients $B_{non-rad}$ are $1.5 \times 10^{-11}$ cm$^3$/s and $4.6 \times 10^{-12}$ cm$^3$/s. According to Eq. (17), the effective $\tau_{band-band}^{non-rad}$ should be 66.8 μs and 215.8 μs if the real density of photo-excited carriers is $10^{15}$ cm$^{-3}$. Besides, Syzgantseva *et al.* predicted the carrier lifetime of pristine UiO-66-NH$_2$ to be 37 ns from NAMD simulation[33]. With $\Delta n(0)=10^{20}$ cm$^{-3}$ used in their simulation, the coefficient $B_{non-rad}$ can be calculated as $2.7 \times 10^{-13}$ cm$^3$/s. If the real density of photo-excited carriers is $10^{15}$ cm$^{-3}$, the effective $\tau_{band-band}^{non-rad}$ should be 3.7 ms instead. As we can see in the two examples, the effective $\tau_{band-band}^{non-rad}$ in the real semiconductors are all very long and much longer than the values derived directly from the NAMD simulations, so it is highly necessary to perform the conversion based on Eqs. (21) and (17) to get the effective $\tau_{band-band}^{non-rad}$. Furthermore, such large values of effective $\tau_{band-band}^{non-rad}$ are consistent it not being a limiting process in determining the total lifetime, so it is necessary to consider other mechanisms and compare $\tau_{band-band}^{rad}$, $\tau_{SRH}$ and $\tau_{Auger}$ to $\tau_{band-band}^{non-rad}$.

According to Eq. (17), the densities of equilibrium carriers, $n_0$ and $p_0$, also influence $\tau_{band-band}$ (similarly for $\tau_{band-band}^{non-rad}$). In Fig. 3a, the dependence of $\tau_{band-band}^{non-rad}$ on $\Delta n(0)$ for three different set of $n_0$ ($p_0$) are considered. In Fig. 3b, the dependence of $\tau_{band-band}^{non-rad}$ on $n_0$ ($p_0$) are explicitly plotted for low and high $\Delta n(0)$. As we can see, when $n_0$ and $p_0$ are both much lower than $\Delta n(0)$, $\tau_{band-band}^{non-rad}$ is almost independent of $n_0$ and $p_0$, because Eq. (17) is simplified into $\tau_{band-band}^{non-rad} = \frac{1}{B_{non-rad}\Delta n(0)}$. This is the case for most of intrinsic semiconductors with the Fermi level close to the middle of band gap, whose $n_0$ and $p_0$ are usually low (for intrinsic CH$_3$NH$_3$PbI$_3$, $n_0$ and $p_0$ can be estimated to be around $10^6$ cm$^{-3}$, much lower than the normal $\Delta n(0)$ around $10^{15}$ cm$^{-3}$ under sunlight). However, when the semiconductors are doped or there are high densities of defects, $n_0$ and $p_0$ can be higher and comparable to $\Delta n(0)$. Here they can reduce the lifetime more significantly, as shown by the shifts of lines in Fig. 3a and the decrease of $\tau_{band-band}^{non-rad}$ at the two ends of Fig. 3b. This can also be seen in Fig. 3c and 3d, in which the simulated decay of $\Delta n(t)$ are obviously faster when $n_0$ is very high or very low (corresponding to very high $p_0$). In the previous supercell simulations, very large $\Delta n(0)$ was assumed, so $n_0$ and $p_0$ were very small compared to $\Delta n(0)$ and their influences can be neglected. However, the real $\Delta n(0)$ in the working photovoltaic devices is only around $10^{15}$ cm$^{-3}$. Doping or intrinsic defects can produce equilibrium carriers with $n_0$ or $p_0$ as high as $10^{17}$ cm$^{-3}$ in CH$_3$NH$_3$PbI$_3$[67,68], then the reduction of $\tau_{band-band}^{non-rad}$ caused by $n_0$ and $p_0$ can be dramatic. In such cases, it is necessary to consider the influences of $n_0$ and $p_0$ when discussing the lifetime of photo-excited carriers.

**Dependence of $\tau_{SRH}$ on Density of Recombination-Center Defects.**

When there are point defects or dopants in a crystal, they not only produce equilibrium carriers (influencing $n_0$ and $p_0$), but also induce the SRH non-radiative recombination of non-equilibrium carriers, thus affecting the lifetime. Eqs. (6) and (7) show the contribution of the SRH recombination to the recombination rate (by $U_{SRH}$) and lifetime (by $\tau_{SRH}$).

In Fig. 4, we simulate the decay of the non-equilibrium carrier density $\Delta n(t)$ according to $\frac{d\Delta n(t)}{dt} = -U_{SRH}$ and Eq. (11) for CH$_3$NH$_3$PbI$_3$ containing recombination-center defects. We assume the electron capture coefficient $c_n$ and hole capture coefficient $c_p$ of the defect in Eq. (14) to be $10^{-7}$ cm$^3$·s$^{-1}$, which are the normal carrier capture coefficients of defects in semiconductors[60,69-71]. Then the SRH recombination coefficient $A$ can be derived at a given defect density $N_t$ following Eq. (14). In Fig. 4a and 4b, the simulated decay of carrier density is shown for the cases when $n_0=p_0=10^6$ cm$^{-3}$, and the defect density $N_t=2.6 \times 10^{20}$ cm$^{-3}$ (corresponding to the high defect density in the NAMD simulations where a defect is put in a supercell with hundreds of atoms) and $N_t = 10^{15}$ cm$^{-3}$ (corresponding to the common defect or doping density in semiconductors), respectively. The decay is fast for high $N_t$ in Fig. 4a, with a short lifetime $\tau_{SRH}=7.8 \times 10^{-5}$ ns. In contrast, $\tau_{SRH}$ becomes much longer and increases to 20 ns when $N_t$ is $10^{15}$ cm$^{-3}$. The comparison shows clearly that $N_t$ has significant influence on $\tau_{SRH}$, as described by Eqs. (14) and (16), which is easy to understand because a higher density of defects can capture more carriers per unit time.

Considering the dependence of $\tau_{SRH}$ on $N_t$, the derived lifetime from the NAMD simulations[72,73] with one defect in a several-hundred-atom supercell ($N_t \approx 10^{20}$ cm$^{-3}$) should not be interpreted as a true lifetime. An approach for extracting the SRH lifetime from the small-supercell NAMD simulations is by converting the value to the $N_t$-independent electron capture coefficient $c_n$ and hole capture coefficient $c_p$ of the defect according to Eqs. (14) and (16), then the lifetime in real samples with different $N_t$ can be calculated using the same equations.

In another example, Shi et al. carried out the simulations of the charge trapping processes in CH$_3$NH$_3$PbBr$_3$ containing the DY$^-$ center[72]. The obtained decay time is 1.9 ns for CBM-to-defect trapping and $4.7 \times 10^{-2}$ ns for VBM-to-defect trapping, respectively. With the defect density ($1.8 \times 10^{20}$ cm$^{-3}$) in their NAMD simulation, we can extract the electron capture coefficient ($c_n$) of $2.9 \times 10^{-12}$ cm$^3 \cdot$s$^{-1}$ and hole capture coefficient ($c_p$) of $1.2 \times 10^{-10}$ cm$^3 \cdot$s$^{-1}$ according to $\tau_{SRH} = \frac{1}{N_t c_n}$ for electron capture and $\frac{1}{N_t c_p}$ for hole capture. The values are low compared to the normal carrier capture coefficients of defects in semiconductors ($10^{-7}$ cm$^3 \cdot$s$^{-1}$)[60,69-71]. Assuming $N_t$ in real semiconductors is $10^{15}$ cm$^{-3}$ and $\Delta n(0) = 10^{15}$ cm$^{-3}$, $\tau_{SRH}$ is 345 $\mu s$ according to Eqs. (14) and (16). Therefore, such a defect should not cause serious limit to the lifetime of photo-excited carriers in MAPbBr$_3$ when the densities of the defect and photo-excited carriers are at the medium level.

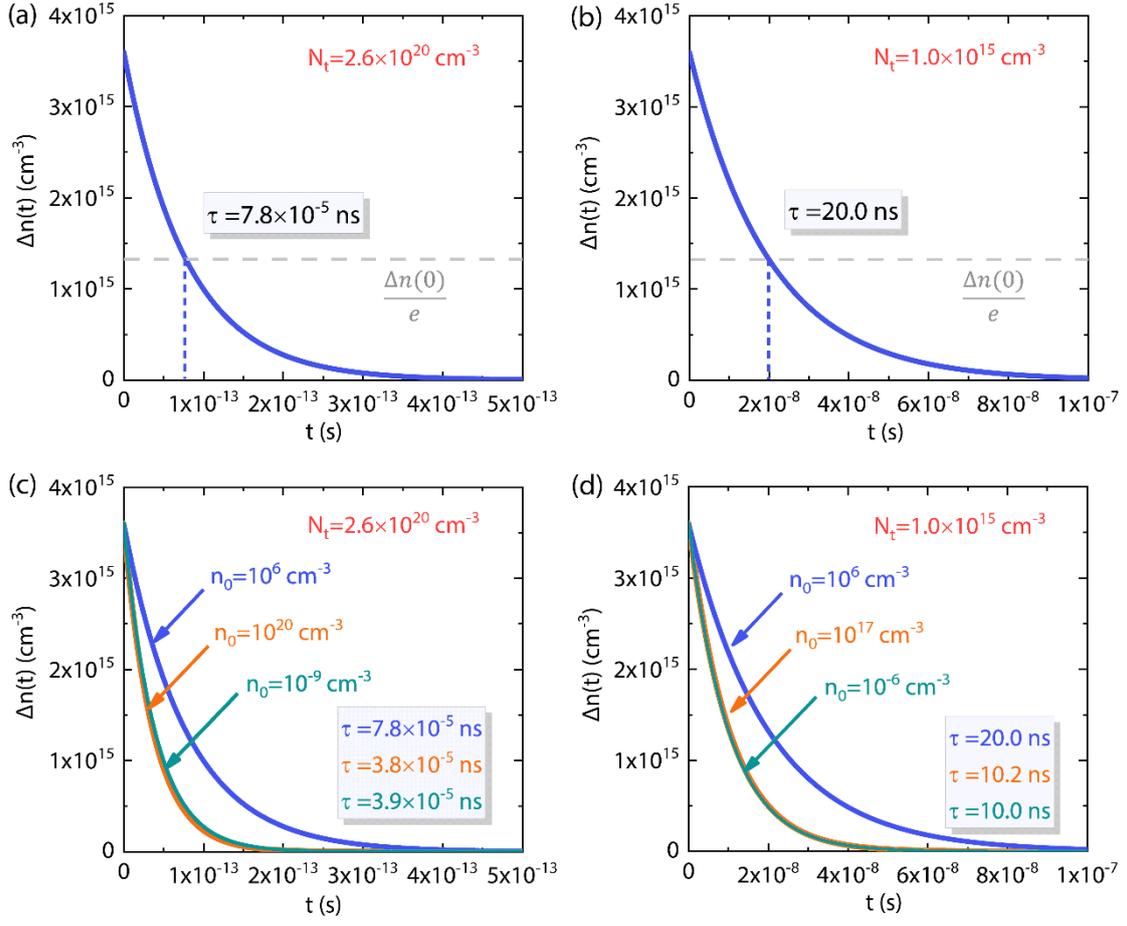

Figure 4. Decay of non-equilibrium carrier density when only the SRH recombination is considered. The non-equilibrium carrier density at t=0 is set to $3.6 \times 10^{15}$ cm$^{-3}$. $N_t$ is $2.6 \times 10^{20}$ $cm^{-3}$ in (a) and (c) and $1.0 \times 10^{15}$ cm$^{-3}$ in (b) and (d). The equilibrium carrier density is $n_0=p_0=0.85\times 10^6 \approx 10^6$ cm$^{-3}$ in both (a) and (b). In (c), $n_0$ varies from $10^{20}$ to $10^{-9}$ cm$^{-3}$ ($p_0$ from $7.3 \times 10^{-9}$ to $7.3 \times 10^{20}$ cm$^{-3}$). In (d), $n_0$ varies from $10^{17}$ to $10^{-6}$ cm$^{-3}$ ($p_0$ from $7.3 \times 10^{-6}$ to $7.3 \times 10^{17}$ cm$^{-3}$).

In Fig. 4a and 4b, only the influence of $N_t$ is shown while $n_0$ and $p_0$ are fixed. However, according to Eqs. (14) and (16), the SRH recombination coefficient $A$ and $\tau_{SRH}$ also depend on $n_0$ and $p_0$. In Fig. 4c and 4d, the influences of $n_0$ and $p_0$ are also simulated. In Fig. 4c different $n_0$ and $p_0$ are considered for high $N_t=10^{20}$ cm$^{-3}$: (i) the sample is highly n-type with a high $n_0=10^{20}$ cm$^{-3}$ and low $p_0= 7.3 \times 10^{-9}$ cm$^{-3}$, (ii) the sample is intrinsic with low $n_0=p_0=10^6$ cm$^{-3}$, and (iii) the sample is highly p-type with a high $p_0=7.3 \times 10^{20}$ cm$^{-3}$ and low $n_0=10^{-9}$ cm$^{-3}$. The simulation shows that $\tau_{SRH}$ becomes shorter in the highly n-type or p-type samples, so the increase of the equilibrium carrier densities can facilitate the recombination. However, the values are still on the same order of magnitude, so the change is small despite the large changes in $n_0$ and $p_0$. When $N_t$ is not very high ($10^{15}$ cm$^{-3}$), the simulation in Fig. 4d also shows that the change of $n_0$ or $p_0$ by $10^{17}$ cm$^{-3}$ induces only small changes in $\tau_{SRH}$, from around 20 ns to 10 ns. From Figs. 4c and 4d, we can see that $\tau_{SRH}$ is mainly determined by $N_t$, and the influences of $n_0$ and $p_0$ are small. This can be understood according to Eq. (14) in which $n_0$ and $p_0$ are present in both the numerator and denominator, so their changes are cancelled largely, giving rise to small influences on the recombination coefficient $A$ and the lifetime $\tau_{SRH}$.

**Comparison of Different Recombination Mechanisms in CH$_3$NH$_3$PbI$_3$.**

The effective lifetime $\tau$ of non-equilibrium carriers is influenced by all recombination mechanisms. Our analysis has shown that $\tau_{SRH}$, $\tau_{band-band}^{non-rad}$, $\tau_{band-band}^{rad}$ and $\tau_{Auger}$ depend on the density of non-equilibrium carriers $\Delta n(0)$, the density of equilibrium carriers $n_0$ and $p_0$, and the density of recombination-center defects $N_t$. When calculating $\tau$, these quantities should be obtained as summarized by the calculation procedure in Fig. 1b. For a semiconductor sample, the density of recombination-center defect or dopant ($N_t$) and the density of the equilibrium carriers ($n_0$ and $p_0$) are usually fixed by the synthesis or annealing conditions. However, the density of non-equilibrium carriers $\Delta n(0)$ in the real samples depends on the environment, *e.g.*, the intensity of the light illumination. To calculate the total effective lifetime $\tau$, *e.g.*, in CH$_3$NH$_3$PbI$_3$ samples, we will first show how to calculate $\Delta n(0)$ under illumination.

In the dark, the density of non-equilibrium carriers is 0. If the above bandgap illumination starts at t = 0, the density $\Delta n(t)$ increases with the rate $\frac{d\Delta n(t)}{dt}$ given by Eq. (1). After a certain time, the system will reach a steady state because the carrier generation and recombination is balanced, and the carrier density $\Delta n(t)$ will maintain at a steady value, as given by

$$\frac{d\Delta n(t)}{dt} = G(t) - U(t) = 0 \quad (22).$$

In Fig. 5a, we simulate the increasing process of $\Delta n(t)$ in the CH$_3$NH$_3$PbI$_3$ thin film.

For the solar illumination on thin films with a layer thickness $d$, the generation rate $G(t)$ of non-equilibrium carriers is a constant and can be calculated as[51],

$$G(t) = (hc)^{-1} \int f_{solar}(\lambda)\alpha(\lambda)\exp(-\alpha(\lambda)d)\lambda d\lambda \quad (23)$$

where $d$ is the film thickness of the absorption layer, $h$ is the Planck constant, $c$ is the speed of light, $f_{solar}(\lambda)$ is the ASTM G173-03 Global Tilt reference spectrum for the solar spectral irradiance distribution, $\alpha(\lambda)$ is the absorbance spectrum of the semiconductor and $\lambda$ is the wavelength of the light. Using the absorption spectrum taken from Ref. [74], the calculated generation rate $G$ is $1.3 \times 10^{21}$ cm$^{-3}$·s$^{-1}$ for the 300 nm thick film. The recombination rate $U(t)$ can be calculated using,

$$U = U_{SRH} + U_{band-band} + U_{Auger}$$
$$= A \cdot \Delta n(t) + (B_{rad} + B_{non-rad}) \cdot [n_0 + p_0 + \Delta n(t)] \cdot \Delta n(t) + C \cdot \Delta n(t)^3 \quad (24),$$

in which the SRH, band-to-band (radiative and non-radiative) recombination and Auger recombination are all considered. The SRH recombination coefficient $A$ is calculated assuming the density of recombination-center defect has two representative values $N_t = 10^{15}$ and $10^8$ cm$^{-3}$, electron capture coefficient and hole capture coefficient $c_n = c_p = 10^{-7}$ cm$^3$·s$^{-1}$ (normal carrier capture coefficients of defects in semiconductors[60,69-71]) and $n_0 = p_0 = 0.85 \times 10^6$ cm$^{-3}$. The band-to-band radiative recombination coefficient $B_{rad} = 1.1 \times 10^{-10}$ cm$^3$/s in CH$_3$NH$_3$PbI$_3$ had been calculated by Zhang et al.[64]. The band-to-band non-radiative recombination coefficient $B_{non-rad} = 2.6 \times 10^{-12}$ cm$^3$/s had been derived above based on the NAMD simulation of Qiao et al.[43]. With $G$ and $U$, $\Delta n(t)$ as a function of $t$ can be simulated, as shown in Fig. 5a.

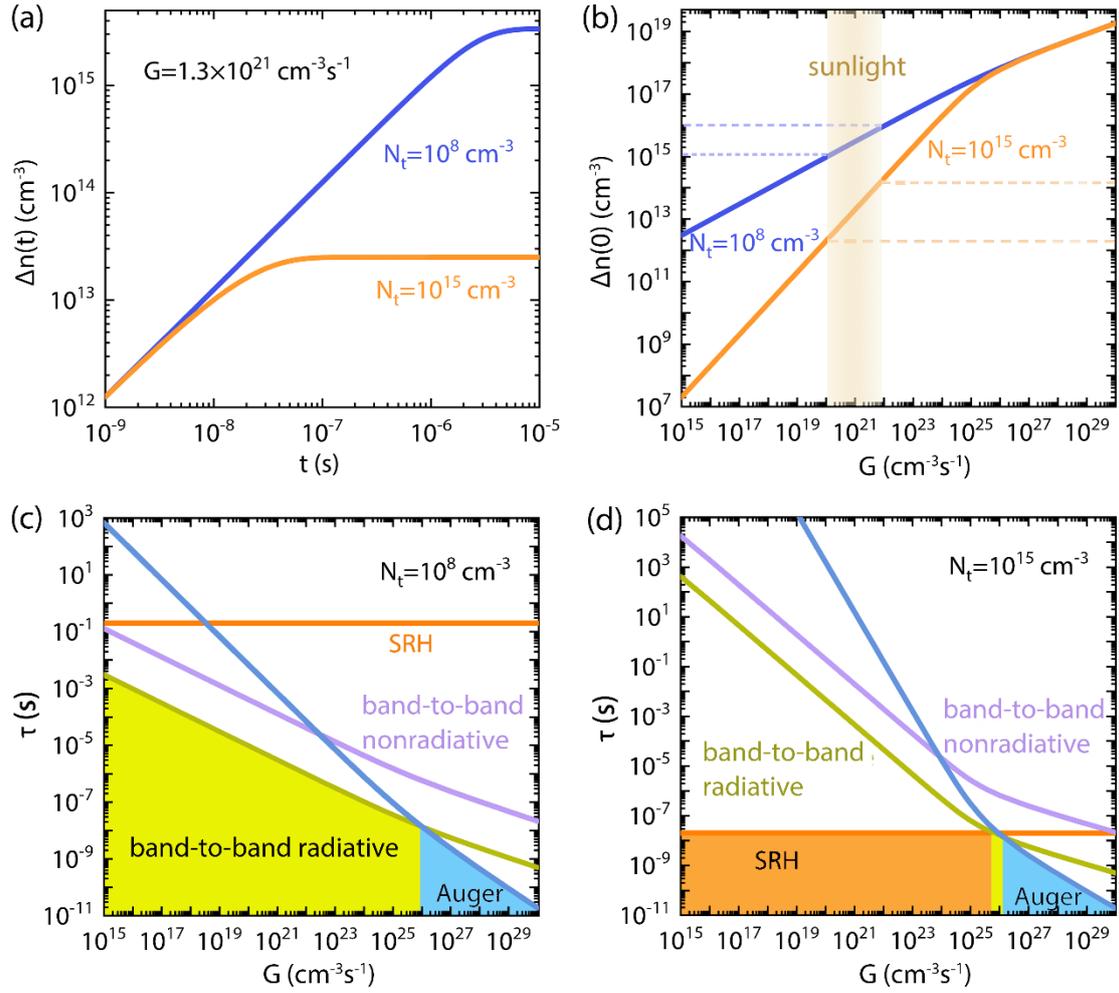

Figure 5. (a) Time evolution of non-equilibrium carrier density under illumination with generation rate $G = 1.3 \times 10^{21}$ cm$^{-3}$·s$^{-1}$ and all recombination mechanisms considered. (b) The steady-state density of non-equilibrium carriers as a function of generation rate. The shaded area shows the range of the generation rates under sunlight illumination. (c) Non-equilibrium carrier lifetime as a function of generation rate with $N_t = 10^8\ cm^{-3}$. (d) Non-equilibrium carrier lifetime as a function of generation rate with $N_t = 10^{15}\ cm^{-3}$. The shaded areas in (c) and (d) show the ranges of generation rate with different dominant recombination mechanisms.

At $t$=0, $\Delta n(t)$ is 0, then it starts to increase under the light illumination. As $t$ increases, the system finally reaches the steady state and $\Delta n(t)$ plateaus. As shown in Fig. 5a, $\Delta n(t)$ reaches a steady value of $2.5 \times 10^{13}\ cm^{-3}$ at $t \approx 10^{-7}$ s when $N_t$=$10^{15}\ cm^{-3}$, while it reaches a steady value of $3.3 \times 10^{15}\ cm^{-3}$ at $t \approx 10^{-5}$ s for $N_t = 10^8\ cm^{-3}$. Therefore, the steady density of non-equilibrium carriers and the time taken to reach steady state are sensitive to the defect density. A higher defect density facilitates SRH recombination and thus decreases the time to reach the steady state. In Fig. 5b, we also simulated the steady density of non-equilibrium carriers for varied generation rate $G$, which corresponds to different illumination intensity. When $G$ is below $10^{26}$ cm$^{-3}$·s$^{-1}$, the steady-state density increases almost linearly as $G$ increases. The generation rate under the sunlight illumination is shown by the shaded area, and the steady density is around $10^{13}$ cm$^{-3}$ for $N_t = 10^{15}$ cm$^{-3}$ and $10^{15}$ cm$^{-3}$ for $N_t = 10^8$ cm$^{-3}$, which are much lower than the density of non-equilibrium carriers (around $10^{20}$ cm$^{-3}$) in small supercells. Only under concentrated light of $10^5$ suns, can the values reach $10^{20}$ cm$^{-3}$.

For a CH$_3$NH$_3$PbI$_3$ thin film with given $N_t$, $E_t$, $c_n$, $c_p$, $n_0$ and $p_0$, the steady density of non-equilibrium carriers under the sunlight illumination should be taken as $\Delta n(0)$ when calculating the lifetime of non-equilibrium carriers in solar cells. Since the steady-state density changes with the generation rate $G$, the lifetime should also change with $G$. When $\Delta n(0)$, $N_t$, $E_t$, $c_n$, $c_p$, $n_0$, $p_0$, $B_{rad}$ and $B_{non-rad}$ are known, $\tau_{SRH}$, $\tau_{band-band}^{rad}$, $\tau_{band-band}^{non-rad}$ and $\tau_{Auger}$ can be calculated directly, as shown in Fig. 1b. Figs. 5c and 5d show how $G$ influences the calculated $\tau_{SRH}$, $\tau_{band-band}^{rad}$, $\tau_{band-band}^{non-rad}$ and $\tau_{Auger}$.

In Fig. 5c, the density of recombination-center defects $N_t$ is set at a low level $N_t = 10^8$ cm$^{-3}$, because most of the deep-level defects in CH$_3$NH$_3$PbI$_3$ have high formation energies and thus low densities[75-79]. The results showed that $\tau_{SRH}$ is always around 0.1 s, which is almost independent of $G$ because the influences of $\Delta n(0)$ on the SRH recombination coefficient A are cancelled partially in the numerator and denominator of Eq. (14). In contrast, $\tau_{band-band}^{rad}$ and $\tau_{band-band}^{non-rad}$ are sensitive to $G$ and decrease quickly as the light intensity and $G$ increase, because the band-to-band radiative and non-radiative recombination coefficients $B_{rad}$ and $B_{non-rad}$ depend on $\Delta n(0)$ almost linearly and thus also on $G$. Comparing $\tau_{SRH}$, $\tau_{band-band}^{rad}$, $\tau_{band-band}^{non-rad}$ and $\tau_{Auger}$, band-to-band radiative recombination is the fastest recombination mechanism and determines the total effective $\tau$ when $G$ is lower than $10^{26}$ cm$^{-3}$·s$^{-1}$, and the Auger recombination dominates at very high generation rates (corresponding to high carrier densities). In contrast, the SRH and band-to-band non-radiative mechanisms only cause slower recombination and thus just decrease $\tau$ slightly according to Eqs. (7) and (20).

For one-sun illumination, $G$ is around $10^{21}$ cm$^{-3}\cdot$s$^{-1}$, and the steady-state density of non-equilibrium carriers is around $10^{15}$ cm$^{-3}$, so the total effective lifetime $\tau \approx \tau_{band-band}^{rad} \approx$ 1 μs. The dominance of the band-to-band radiative recombination and the predicted long effective lifetime $\tau$ are consistent with the experimental finding of efficient photoluminescence and measured long carrier lifetime around 1 μs in the CH$_3$NH$_3$PbI$_3$ thin films[80].

If the density of recombination-center defects $N_t$ is increased, the dominant recombination mechanism can be changed. Fig. 5d shows the case for higher $N_t = 10^{15}$ cm$^{-3}$, which is common for the deep-level defects in many semiconductors. $\tau_{SRH}$ is now shorter, around 10 ns. The SRH recombination determines the total effective lifetime $\tau$ when the light intensity is low and $G$ is lower than $10^{25}$ cm$^{-3}\cdot$s$^{-1}$, while the band-to-band radiative recombination determines $\tau$ only in a small range of G around $10^{26}$ cm$^{-3}\cdot$s$^{-1}$. When the light intensity is high and $G$ is higher, Auger recombination becomes dominant. Therefore, there is a transition of the dominant recombination mechanism as the light intensity and $G$ increase. For the sunlight illumination, the steady-state density of non-equilibrium carriers is around $10^{13}$ cm$^{-3}$ (Fig. 5b), so SRH recombination is dominant.

Our analysis confirms that the lifetime of non-equilibrium carriers in CH$_3$NH$_3$PbI$_3$ solar cells should be mainly determined by the band-to-band radiative recombination and is usually very long. Only when thin films have high density of recombination-center defects or dopants, can SRH recombination appreciably decrease the lifetime. Across whole range of $G$, $\tau_{band-band}^{non-rad}$ is always longer than $\tau_{band-band}^{rad}$ by two orders of magnitude, so the influence of band-to-band non-radiative recombination on the effective $\tau$ is negligible in the CH$_3$NH$_3$PbI$_3$ thin films, no matter under strong or weak light illumination. Therefore, changes in $\tau_{band-band}^{non-rad}$ alone should not be overinterpreted, *e.g.*, 1.5 ns for the pristine CH$_3$NH$_3$PbI$_3$ compared to 4-11 ns for doped or defective crystals[43]. The experimentally improved performance of CH$_3$NH$_3$PbI$_3$ solar cells after the alkaline metal treatments should result from the significantly increased $\tau_{SRH}$, because the alkaline metal dopants passivate the recombination-center level of the I interstitial and thus give a clean band gap.

**Concluding Remarks**

This study is motivated by the development of more quantitative predictions of carrier recombination processes and the lifetimes of non-equilibrium carriers in semiconductors. From the recent literature, there is a striking discrepancy between the carrier lifetimes derived directly from the NAMD simulations using small supercells under periodic boundary conditions and the experimentally measured carrier lifetimes.

By revisiting the fundamental definition of carrier lifetime and considering four recombination mechanisms, we were able to develop a systematic procedure for calculating the effective lifetime of non-equilibrium carriers in semiconductors. Within this procedure, the NAMD and other methods are combined to calculate the effective lifetime that can be compared directly to the experimental values. The calculated results reinforce fundamental concepts in the field, and demonstrate that two factors have significant influence on calculated lifetimes: (i) the consideration of radiative and non-radiative recombination mechanisms, and (ii) the density of equilibrium carriers, non-equilibrium carriers, and recombination-center defects. Recent calculations of carrier lifetimes in $CH_3NH_3PbI_3$ and other optoelectronic semiconductors were limited by (i) the neglect of band-to-band radiative recombination, and (ii) an exaggerated density of non-equilibrium carriers and defects due to the small supercells, which cause the obvious discrepancies between calculations and experiments. Using our procedure, these discrepancies are overcome and the calculated effective lifetime of $CH_3NH_3PbI_3$ can agree well with experiments. Furthermore, our analysis revealed that the effective lifetime is determined by band-to-band radiative and defect-assisted non-radiative recombination, rather than the band-to-band non-radiative recombination ($\tau_{band-band}^{non-rad}$) whose influence is always negligible. Previous conclusions based on the interpretation of changes in $\tau_{band-band}^{non-rad}$ should be revisited. These results demonstrate that it is possible to calculate the carrier lifetime in agreement with experimentally measured values based on the NAMD simulations. The systematic procedure reported here can enable more accurate future studies of carrier dynamics in semiconductors.

**Acknowledgements**

This work was supported by Shanghai Academic/Technology Research Leader (19XD1421300), National Natural Science Foundation of China (NSFC) under grant Nos. 12174060, 11991060, 12088101 and U1930402, Program for Professor of Special Appointment (Eastern Scholar TP2019019), National Key Research and Development Program of China (2019YFE0118100), State Key Laboratory of ASIC & System (2021MS006) and Young Scientist Project of MOE Innovation Platform. X. G. was supported by NSFC under grant No. 12188101.